\documentclass[nopreprintline]{elsarticle}

\usepackage{lineno}
\modulolinenumbers[5]

\usepackage{amssymb}
\usepackage{url}
\usepackage{amsfonts,amsmath}
\usepackage{graphicx}
\usepackage{stfloats}
\usepackage{subfigure}

\usepackage{multirow}
\usepackage{xfrac}
\usepackage{color}
\usepackage{array}
\usepackage[table]{xcolor}

\usepackage[Symbol]{upgreek}

\usepackage[T1]{fontenc}

\usepackage{graphicx,url}
\usepackage{picinpar}
\usepackage{amsmath}
\usepackage{stfloats}
\usepackage{url}
\usepackage{flushend}
\usepackage[latin1]{inputenc}
\usepackage{colortbl}
\usepackage{soul}
\usepackage{multirow}
\usepackage{pifont}
\usepackage{color}
\usepackage{alltt}
\usepackage[hidelinks]{hyperref}
\usepackage{enumerate}
\usepackage{siunitx}
\usepackage{breakurl}
\usepackage{epstopdf}
\usepackage{pbox}
\usepackage{algorithm}
\usepackage{algpseudocode}

\begin{document}

\begin{frontmatter}

\title{High-Performance Parallel Implementation of Genetic Algorithm on FPGA}

\author{Matheus F. Torquato and Marcelo A. C. Fernandes}
\address{Department of Computer Engineering and Automation \\ 
Research group on Embedded System and Reconfigurable Computing (RESRC) \\
Federal University of Rio Grande do Norte (UFRN)\\Natal, Brazil}
\ead{mfernandes@dca.ufrn.br}




\begin{abstract}
Genetic Algorithms (GAs) are used to solve search and optimization problems in which an optimal solution can be found using an iterative process with probabilistic and non-deterministic transitions. However, depending on the problem's nature, the time required to find a solution can be high in sequential machines due to the computational complexity of genetic algorithms. This work proposes a parallel implementation of a genetic algorithm on field-programmable gate array (FPGA). Optimization of the system's processing time is the main goal of this project. Results associated with the processing time and area occupancy (on FPGA) for various population sizes are analyzed. Studies concerning the accuracy of the GA response for the optimization of two variables functions were also evaluated for the hardware implementation. However, the high-performance implementation proposes in this paper is able to work with more variable from some adjustments on hardware architecture.
\end{abstract}


\end{frontmatter}


\section{Introduction}\label{sec:intro}

In the last years, the increasing number of critical applications involving real time systems in conjunction with the growth of integrated circuits density and the continuous reduction in the power supply voltages transformed the development of new suitable computational solutions an even harder task to achieve. Due to the intense demand in the electronics goods market for high processing speeds at smaller time frames, without neglecting the energy savings, the technology industry has faced an extremely competitive and challenging scenario in terms of designing hardware solutions to meet this constantly growing demand. One way found by researchers and developers to address such demands is by using algorithm parallelization techniques. Parallel processing is used to manipulate data concurrently, so that while computing one section of the algorithm, other stations perform similar operations on another set of data \citep{[10]}. Combining the hardware implementation with the parallelization of algorithms is often a satisfactory solution for high performance and higher speed applications when compared to sequential solutions.

The Field Programmable Gate Arrays (FPGAs) are reconfigurable hardware devices suited to this scenario due to the nature of its architecture. Given that FPGAs are huge configurable gates, they can be programmed to operate as multiple parallel paths in hardware. In this way, there is a real parallelization in which the running operations do not need to compete for the same resources since each one will be executed by different gates \citep{[22]}. The increasing density and price reduction of FPGAs expand the opportunities for developers and researchers to use higher density FPGA devices for hardware implementations \citep{[13]} considering the use of such devices is advantageous since the development time and costs are significantly reduced \citep{[19]}.

The convergence among genetic algorithms, parallelization techniques and reconfigurable hardware implementation results in this work which presents a proposal of parallel implementation of a genetic algorithm on FPGA. This paper focuses on high-performance and critical applications that require nanoseconds time constraints to be satisfied. On the other hand, in applications where processing speed is not the critical factor or it is less limiting than the necessity for low power consumption, it is possible to decrease the energy utilization by reducing the clock cycles rate, considering that the dynamic power utilization is diminished when an operating frequency lower than the maximum theoretical one is used \citep{[46]}. Applications that process a large flow of data can be benefited and accelerated by this implementation here developed. Some applications examples are: data mining, tactile internet, massive data processing and bioinformatics.

\subsection{Related Work}\label{Related_work_section}

Genetic algorithms and Artificial Intelligence (AI) have long been used in applications of the most diverse areas to optimize and find satisfactory solutions in computing, engineering and other fields. More recently, a wider range of applications and variations of genetic algorithms such as parallel and distributed applications, hardware implementations, new proposals for genetic operators, and hybrid (software and hardware) implementations  of genetic algorithms have been observed within the research scenario.

In \citep{[5]}, it is proposed an implementation of a customizable Intellectual Property Core (IP Core) for FPGA that implemented a general-purpose genetic algorithm. In this work, the authors have focused on the genetic algorithm programmability implemented in the IP core. The customization could be done regarding population size, generation number, crossover and mutation rates, random number generators seeds and the fitness function. One of the work's highlights is the support for a multiple of these functions. The proposed IP can be programmed with up to eight fitness functions which could be synthesized in conjunction with the GA and implemented in the same FPGA device. The proposed core also has additional input/output ports that allow the user to add further fitness functions that have been implemented on a second FPGA device or some other external device. The implementation utilized $13\%$ of the available logical cells of a Xilinx Virtex II Pro (xc2vp30-7ff896). However, since a trade-off between performance and flexibility exists, and once the authors focused on flexibility over performance, the speedup over analogous software implementation was only of $\times5.16$.

Hardware Genetic Algorithms implementations can also be observed in \citep{[6]}, \citep{[9]}, \citep{[14]} and \citep{[20]}. The work detailed in \citep{[20]} showed the OIMGA which its strategy was to retain only the ideal individual from the population making the memory requirements drastically reduced.  The paper \citep{[6]} presented a compact implementation of a genetic algorithm on FPGA that represented the population of chromosomes as a vector of probabilities. The work focused on the lower consumption of memory, power and space resources in hardware, but it was not fully implemented on FPGA as it used a software written in C++ to compute the values from the fitness function. The work \citep{[14]} proposed a high-speed GA implementation on FPGA. The implementation was based on the HGA proposed by \citep{[30]}, the first known GA implementation on FPGA, and the authors claimed that the developed system surpassed any existing or proposed solution according to their experiments. The P-HGAv1, version developed by \citep{[14]} of the HGA claimed to be parametric, have low silicon requirements and support multiple fitness functions. Although the authors have focused on the speed of the algorithm and reached a time of $0.021$ milliseconds for each GA generation, this speed may not be compatible with real-time applications that require low latency.

The works presented by \citep{[4]}, \citep{[8]} and \citep{[17]} showed applications for ground mobile robots using GAs, where these first two were embedded implementations on FPGA. \citep{[4]} developed, according to the authors, the first GA-based hardware implementation of a simultaneous localization and mapping (SLAM) system for ground robots. The authors achieved significant hardware acceleration compared to software implementation by exploiting the pipelining and parallelization capabilities of reconfigurable hardware. In this project the GA's genes that made up the population represented possible robot movements based on the previous position. Later, In the work developed by \citep{[8]}, the goal was to determine the optimal movements considering various aspects such as route tracking and low energy consumption, avoiding obstacles collision. The authors pointed out that the implementation was suitable for real-time use and stated that all GA stages have been implemented in hardware modules. The solution presented in this work offered a convergence time of less than $2$ milliseconds, it used $17124$ out of the $17600$ ($97$\%) Lookup tables (LUTs) available in the FPGA, but the frequency obtained after the synthesis process was not informed. \citep{[17]} developed a genetic algorithm with a coevolutionary strategy for global trajectory planning of several mobile robots. According to \citep{[31]}, co-evolution is the process of mutual adaptation of two or more populations simultaneously, and it was used to reflect the fact that all species are simultaneously co-evolving in a given physical environment. The implementation of \citep{[17]} promised an improvement in the genetic operators of conventional GAs and proposed a new operator of genetic modification, but these developments were not implemented in hardware.

The implementations seen in \citep{[12]}, \citep{[7]} and \citep{[2]} were GA  applications in digital signal processing and control systems embedded on FPGA. \citep{[12]} presented a real-time GA for adaptive filtering application with all modules implemented in hardware such as fitness function, selection, crossover, mutation and random number generator functions. The implementation was designed in hardware and after its synthesis, a rate of $320$ thousands generations per second was achieved. Meanwhile, \citep{[7]} proposed a GA for multi-carrier dynamic systems based on filter banks. The authors of \citep{[2]} proposed a design and an implementation of a PID controller based on GA and FPGA. The researchers stated that the design method of the intelligent PID controller based on FPGA and GA was successfully verified and had some advantages such as flexible design, automatic online tuning, high reliability, short technical development cycle and high execution speed. For this case, each GA chromosome was coded with the controller set of gains $K_p$, $K_i$ and $K_d$. Details of FPGA area occupancy and obtained throughput were not reported.

Lastly, \citep{[3]} and \citep{[11]} presented parallel and distributed implementation of GA using FPGAs. \citep{[3]} proposed a solution for parallel genetic algorithms in multiple FPGAs. Using multiple populations in parallel GAs was based on the idea that population isolation can maintain greater genetic diversity, while communication between them can cause GAs to work together to find good solutions. The implementation of \citep{[3]} was applied to three different benchmarks, including the traveling salesman problem, and the authors stated from the experimental results that in a configuration of $4$ FPGAs an average acceleration of $30$ times over a multi-core processor GA was achieved. \citep{[11]} introduced GRATER, an automated design workflow for FPGA accelerators that leveraged imprecise computation to increase data-level parallelism and achieved higher computational throughput. In this work the main idea was establishing a negotiation among circuit that involved area, energy and performance in exchange for precision reduction. This was achieved through an imprecise implementation of specific hardware blocks such as adder and multiplier, since hardware area reduction resulted in better data parallelism utilization and, therefore, increased the yield. Also in \citep{[11]}, genetic programming was used to evolve variants of the input kernel until one was found with ideal assignments which reduced the synthesized kernel area  while still stochastically satisfying  the output desired quality. The synthesis results in an Altera Stratix V FPGA showed that the reduced area of the approximate kernels produced a gain of $\times1.4$ to $3.0$ higher with less than $1\%$ of quality loss compared to benchmarks.

It is essential to observe that the works presented in the literature propose the solutions based on software and hardware on FPGA. This kind of the answer increase de flexibility but decrease the throughput. Thus, Differently of the papers in the literature, this work proposed a high-performance parallel implementation of GA. The implementation uses the full-parallel strategy in all GA operations in order to maximize the throughput.

\subsection{Paper organization}

In section \ref{Section_AG} a theoretical foundation about the meta-heuristic used here will be explored as well as the genetic algorithms, its main characteristics, its advantages and its different applications. Section \ref{Implementation_Description} will present a detailed description of the architecture development and implementation, describing the various hardware modules used to construct the parallel genetic algorithm. Later, in section \ref{Results_sec}, a careful analysis of the results obtained from the implementation described in the previous section will be performed. Simulation results, synthesis in the FPGA and the validation of the proposed architecture will be presented. The analysis will be carried for parameters such as occupation area and sampling frequency, taking into account different configurations of the proposed architecture embedded in the reconfigurable hardware. Following, Section \ref{Comparisons_sec} shows a comparison of the obtained with other similar works found in the state of the art. Finally, Section \ref{conclusion_sec} will present final considerations, conclusions on the results obtained.

\section{Genetic Algorithms}\label{Section_AG}

The GAs are used to solve search and optimization problems where an optimal solution can be found through an iterative process in which the search starts from an initial population and then, combining the best representatives of it, obtains a new population that replaces the previous one \citep{[39]}.

GA is an iterative algorithm that is started from a population of $N$ chromosomes randomly created. $N$ is even, in the case of this proposed work, in order to facilitate the implementation. In every $k$-th iteration, also called generation or epoch, the $N$ chromosomes are evaluated, selected, recombined and mutated to form a new population also of $N$ chromosomes, that is, the entire population of parents is replaced by the new offspring. Then, the new population is used as input to the algorithm's next iteration (generation), and this procedure of population updating is repeated $K$ times, where $K$ is the GA generations number.

The Algorithm \ref{ResumoAG} displays the pseudo code of  a GA. This code details all the variables and procedures that will be used in the implementation to be presented in the following sections. The variable $x_j[m](k)$ represents the $j$-th chromosome of $m$ bits in the $k$-th generation and $\mathbf{X}[m](k)$ is a vector that stores all the $N$ chromosomes, that is,
\begin{equation} \label{EqMatInd}
\mathbf{X}[m](k) = \left[
    \begin{matrix}
    x_1[m](k) \\ 
    \vdots \\
    x_N[m](k)
    \end{matrix}
    \right]
    .
\end{equation}

After the initialization process, the fitness function, called \emph{FF} (Line \ref{CodFA} of Algorithm \ref{ResumoAG}), calculates the fitness value of the $N$ chromosomes $x_j[m](k)$ of the population. This operation is applied to all chromosomes and results in a respective value $y_j[a](k)$ for each $j$-th chromosome, where $b$ is the number of bits representing the fitness value. The better the value $y_j[a](k)$ of the chromosome $x_j[m](k)$, the more likely it is to continue in the new generations. The fitness values of all $N$ individuals are stored in
\begin{equation} \label{EqMatApt}
\mathbf{Y}[a](k) = \left[
    \begin{matrix}
    y_1[a](k) \\ 
    \vdots \\
    y_N[a](k)
    \end{matrix}
    \right].
\end{equation}

\begin{algorithm}[t]
\caption{Genetic Algorithm}
\label{ResumoAG}
\begin{algorithmic}[1]
   \State Initialise $\mathbf{X}[m](k)$ with random values
   \For{$k\gets1$ \textbf{to} $K$}\label{Loop1}
  \For{$j \gets 1$ \textbf{to} $N$}\label{Loop2}
      \State $y_j[a](k) \gets \text{\emph{FF}}\left(x_j[m](k)\right)$ \label{CodFA}
   \EndFor \label{EndLoop2}
   \For{$j\gets1$ \textbf{to} $N$}\label{Loop3}
    \State $w_j[m](k) \gets\text{\emph{SF}}\left(\mathbf{Y}[b](k),\mathbf{X}[m](k)\right)$ \label{CodFS}
    \EndFor \label{EndLoop3}
   \For{$i\gets1$ \textbf{to} $N/2$}\label{Loop4}
    \State $\left[
    \begin{matrix}
    z_{2i-1}[m](k) \\ z_{2i}[m](k)
    \end{matrix}
    \right] \gets \text{\emph{CF}}\left(\left[
     \begin{matrix}
    w_{2i-1}[m](k) \\ w_{2i}[m](k)
    \end{matrix}
   \right]\right)$ \label{CodFC}    
    \EndFor \label{EndLoop4}
    \For{$v\gets1$ \textbf{to} $P$}\label{Loop5}
    \State $x_v[m](k)\gets \text{\emph{MF}}\left(z_v[m](k)\right)$ \label{CodFM}    
     \EndFor \label{EndLoop5} 
   \EndFor \label{EndLoop1}
\end{algorithmic}
\end{algorithm}

After calculating the fitness value of each $j$-th chromosome of the $k$-th generation, the selection operation is performed. In GAs, the selection's purpose is to highlight the chromosome $x_j[m](k)$ alongside its respective fitness values, $y_j[a](k)$, in order to produce better future populations. There is a great variety of selection methods in the literature and among them it can be mentioned: the method of selection by ranking, by tournament, roulette selection and elitism. The tournament selection method used in this implementation is one of the most used \citep{[40]} and it makes a competition between two or more randomly chosen chromosomes from the population stored in $\mathbf{X}[m](K)$. This competition consists of comparing the strength (fitness), $y_j[a](k)$, of all participating chromosomes and the one who holds the best respective value in $\mathbf{Y}[a](K)$,  proceeds in the algorithm to pass their genes forward. The selection function, called here \emph{SF} (Line \ref{CodFS} of the Algorithm \ref{ResumoAG}), has the vectors $\mathbf{Y}[a](k)$ and $\mathbf{X}[m](k)$ from the $k$-th generation as it inputs and, for each input value, it outputs the variable $w_j[m](k)$ that can assume the value of any of the $N$ chromosomes stored in $\mathbf{X}[m](k)$. All $w_j[m](k)$ values are grouped in
\begin{equation} \label{EqMatW}
\mathbf{W}[m](k) = \left[
    \begin{matrix}
    w_1[m](k) \\ 
    \vdots \\
    w_N[m](k)
    \end{matrix}
    \right]
\end{equation}
in order to be used in the crossover stage.

The crossover stage in the $k$-th generation occurs after the selection of the most fit chromosomes in the population (stored in $\mathbf {W} [m] (k) $) by the selection function and aims to originate new chromosomes of which will, after the mutation stage, compose the next GA generation updating the vector $ \mathbf {X} [m] (k) $). There are several crossover techniques presented in the literature and the strategy adopted in this implementation was the single point crossover. The crossover operation, called here \emph {CF} (Line \ref{CodFC} of the Algorithm \ref{ResumoAG}), has as input pairs of elements from vector $ \mathbf {W}[m](k)$ of the $k$-th generation and as output, pairs of 
\begin{equation} \label{EqMatZ}
\mathbf{Z}[m](k) = \left[
    \begin{matrix}
    z_1[m](k) \\ 
    \vdots \\
    z_N[m](k)
    \end{matrix}
    \right]
\end{equation}
which stores the chromosomes after crossingover, that is, the new $k$-th offspring.

The last GA's step is the mutation operation that changes the value of a group $P$  chromosomes, in order to provide greater diversity to the population avoiding its solution to stabilise in local minimums. The mutation rate, $MR$, is the parameter responsible for controlling the amount of mutated chromosomes. Normally, the $MR$ ranges from $ 0.1\% $ to $ 2\% $. The $ P $ value can be easily calculated by the expression
\begin{equation} \label{EqExpP}
P = \lceil N \times MR \rceil.
\end{equation}

The mutation operation, referred here as \emph {MF}, is presented in the Line \ref{CodFM} of the Algorithm \ref{ResumoAG} and detailed in Equation \ref{Operacao_Mutacao}.
\begin{equation} 
\label{Operacao_Mutacao}
x = (\neg z  \land Rand) \lor (z  \land \neg Rand).
\end{equation}
Where $ z $ is the chromosome to be mutated and $ Rand $ is a random variable. The result of this exclusive OR operation between $ z $ and $Rand$ ($Z\oplus Rand$)  results in $ x $.

\section{Hardware Proposed}\label{Implementation_Description}

\begin{figure*}[t!]
  \centering
  \includegraphics[width=11cm,keepaspectratio]{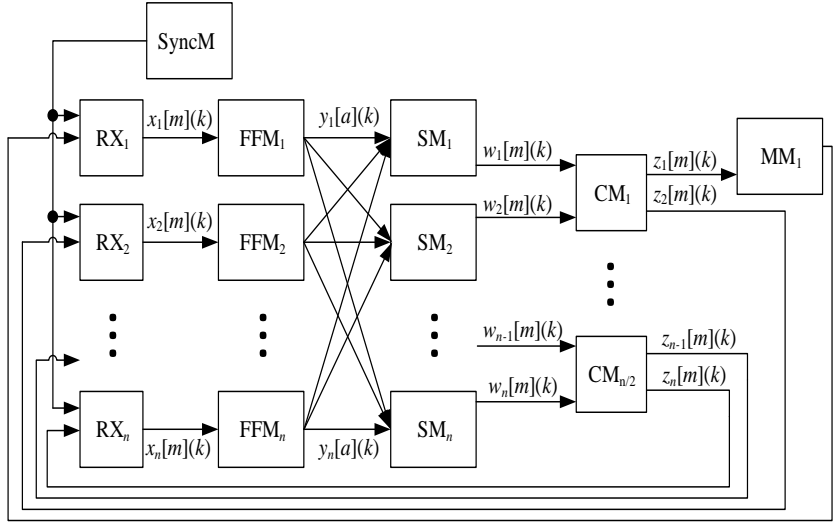}
  \caption{General architecture of the proposed parallel genetic algorithm implementation.}
  \label{AG_Estrutura_Geral}
\end{figure*}

Figure \ref{AG_Estrutura_Geral} presents the general architecture of the proposed GA hardware implementation. The entire algorithm was developed using a parallel architecture focusing on accelerating the processing speed, taking advantage of the available hardware resources, similarly to \citep{[50]}. The Figure details in block diagram the main subsystems of the proposed implementation, which in turn were encapsulated in order to make the general visualization of the architecture less complex. It is possible to observe a population of $ N $ chromosomes of $ m$ bits in which $ x_j [m] (k) $ represents the $j$-th chromosome of the population in the $k$-th generation, according to the Algorithm \ref {ResumoAG}. Each $j$-th chromosome $x_j[m](k)$ is stored in a $m$-bit register, called here $\textrm{RX}j$ whose value is updated by the new population of $ N $ chromosome produced after the processes of selection, crossover and mutation. This updating process occurs every time the synchronization module, called here SyncM, enables the registers to store new values.

Given that the implementation optimizes two-variables functions, each register $\textrm{RX}j$ stores the values of both binary inputs for the fitness function using bits concatenation for such storage. The first $\frac{m}{2}$ bits represent the first input of the fitness function,  $px_j[\frac{m}{2}](k)$, while the second block of $\frac{m}{2}$ bits stores the second input for the fitness function,  $qx_j[\frac{m}{2}](k)$. Thus, 
\begin{equation}
\label{Register_storage}
x_j[m](k) = px_j\left[\frac{m}{2}\right](k) \Vert  qx_j[\frac{m}{2}](k)
\end{equation}
where $\Vert$ is the concatenation operator.

The initial population of the algorithm is randomly chosen. All random values from the present implementation is generated by pseudo random number generators based on Linear Feedback Shift Register (LFSR) \citep{[51]} and \citep{[52]}. $32$ bits independent LFSRs based on the polynomial $ r^{32} + r^{22} + r^{2} + 1 $ \citep{[52]} were used. Each generator is characterised as CCLFSR$lj$ whose CC, $ l $ and $ j $ are labels for its position in the circuit. Every $ k $-th generation a random variable of $ 32 $ bits, called here $\text{CC}r_{lj}[32](k)$, is produced by each LFSR. To avoid the same sequence of values, each generator LFSRCC$lj$  has a different initial value of $ 32 $ bits, called $\text{CCseed}_{lj}[32]$.

The notation used in the following diagrams will be in the $x[m](c)$ form, where $ x $ is the variable, $ m $ is the bit word width and $ c $ represents the generation of the genetic algorithm ranging from 1 to $ K $. In some cases only the bracketed notation, $ [m] $, will be shown to represent the amount of bits transferred on the bus.

The implementation consists of five main modules called: Fitness Function Module (FFM), the Selection Module (SM), Crossover Module (CM), Mutation Module (MM) and Synchronization Module (SyncM). Each module has its specific implementations that will be detailed in the following sections.

\subsection{Fitness Function Module - FFM}\label{FFModule_Section}

\begin{figure*}[b]
  \centering
  \includegraphics[width=12cm,keepaspectratio]{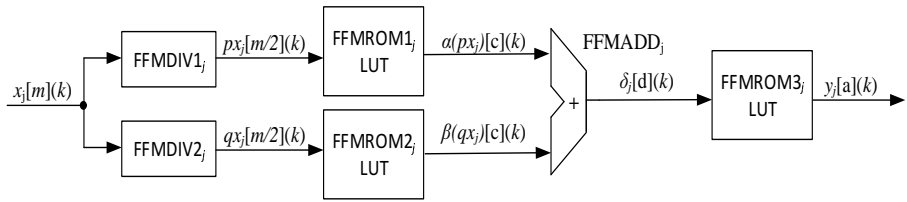}
  \caption{Fitness Function Module - FFM}
  \label{FFM_fig}
\end{figure*}

The Fitness Function Module (Figure \ref{FFM_fig}) has the purpose of calculating the fitness value of each $j$-th chromosome from a fitness function $f(\cdot)$. The proposed structure has $N$ FF modules and each $j$-th module, called here FFM$ _j $, is associated to an individual $x_j[m](k)$ and generates as output in every $k$-th generation a fixed-point fitness value expressed by
\begin{equation}
\label{FF_eq}
y_j[a](k) = \text{FFM}j(x_j[m](k)),
\end{equation}
where $ a $ represents the bit width (equivalent to the \ref {CodFA} line of the Algorithm \ref {ResumoAG}).

Not only for the Fitness Function Module, but for all other stages, the proposed architecture is capable of solving one or two variable problems. Regardless of the case, the user will not need to make any adjustments to the input data. The difference between these two options reflects only on how the data is manipulated by the subsequent modules, but this does not change the performance of the system and is done invisibly to the operator.

Figure \ref{FFM_fig} details the operation of the $j$th FFM. The FFM input value, $x_j[m](k)$ stored in the RX$_j $ register, is divided into two halves of $\frac{m}{2}$ bits, $px_j[\frac{m}{2}](k)$ and $qx_j[\frac{m}{2}](k)$, by the bit splitters FFMDIV1$_j$ and  FFMDIV2$_j$  so that it is thus possible to operate each variable independently in the case of two variables problems.



After split, the variable $px_j[\frac{m}{2}](k)$ is directed to the ROM memory FFMROM1$ _j $ which implements the $\alpha$ function through a Look-Up Table (LUT) and the variable $qx_j[\frac{m}{2}](k)$ is directed to FFMROM2$ _j $ which implements the $\beta $ function in the same fashion.

After this, both values are added by the FFMADD$ _j $ adder resulting in the $\delta_j[d](k)$ variable, where
\begin{equation} \label{Eq_FFM1}
\delta_j[d](k) = \alpha(px_j)[c](k) + \beta(qx_j)[c](k).
\end{equation}
The variable $\delta_j[d](k)$ is then directed to the LUT FFMROM3$_j $ where the $\gamma$ function will be implemented, hence
\begin{equation} \label{Eq_FFM2}
y_j[a](k) = \gamma(\delta_j[d](k)).
\end{equation}

In general, the FFM shown in Figure \ref{FFM_fig} is able to solve any one or two variables problem in the format
\begin{equation} \label{EqGenerica}
y_j[a](k) = \gamma(\alpha(px_j[c](k)) + \beta(qx_j[c](k))).
\end{equation}
Expressions with product between the two variables are not possible in this current approach, but it would be possible through a change in the structure of the FFM.

\subsection{Selection Module - SM}
\label{SM_Section}

\begin{figure*}[h]
  \centering
  \includegraphics[width=12cm,keepaspectratio]{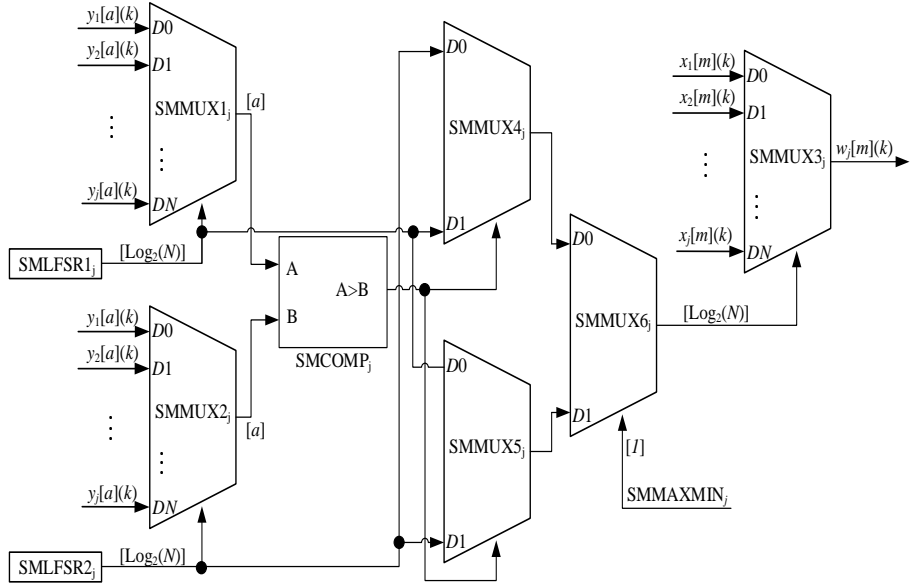}
  \caption{Selection Module - SM}
  \label{SM_fig}
\end{figure*}

The selection module (SM) implements the tournament selection method, as mentioned in Section \ref{Section_AG}, by doing a competition between two chromosomes. Similarly to the FFM, there are $N$ SMs for a group of $N$ chromosomes. As detailed in Figure \ref{SM_fig}, each $j$-th SM, here called SM$j$, has as input the $N$ fitness values, $y_j [a] (k)$, and $N$ chromosomes, $ x_j [m] (k) $, from the $k$-th generation (equivalent to the \ref{CodFS} line of the Algorithm \ref {ResumoAG}).

\begin{figure*}[b!]
\begin{center}
  \includegraphics[scale=0.6]{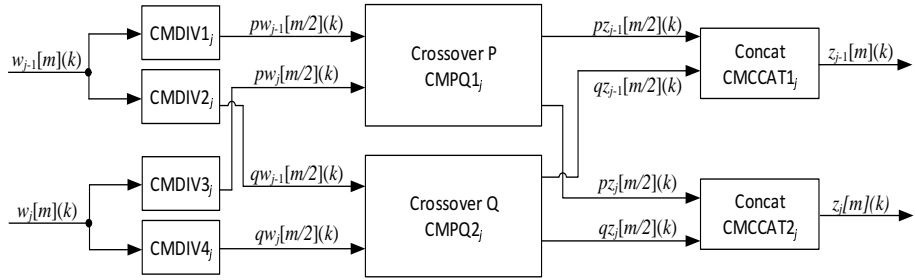}\\
  \caption{$J$-th Crossover Module (CM$_j$)}  
  \label{Cruzamento_esquema}
\end{center}
\end{figure*}

Each $j$-th $ \textrm {SM} $ has two random generators called SMLFSR1$j$ and SMLFSR2$j$. In addition to the random generators, this module is formed by three $N$ input multiplexers called here SMMUX1$j $, SMMUX2$j $ and SMMUX3$j $, a $m$ bits comparator, called SMCOMP$j $ and three two-input multiplexers, called SMMUX4$j $, SMMUX5$j$ and SMMUX6$j$.

The SMMUX1$j$ and SMMUX2$j$ multiplexers are driven by the SMLFSR1$j$ and SMLFSR2$j$  generators output signal, ($\text{SM}r_{1j}[32](k)$ and $\text{SM}r_{2j}[32](k)$), respectively. As shown in Figure \ref{SM_fig} the output signal of each generator ($\text{SM}r_{1j}[32](k)$ and $\text{SM}r_{2j}[32](k)$) is truncated in the most significant $ \left \lceil log_2(N) \right \rceil$ bits in order to match the population size. The SMMUX1$j$ and SMMUX2$j$ multiplexers select one fitness value each, which is related to its correspondent chromosome by its index value.

Finally, SMMUX3$j$ selects the chromosome associated with the best  fitness function value from the output of SMMUX6$j$ which selects whether the goal is to maximize or minimize the evaluation function through the SMMAXMIN$j$ variable.

\subsection{Crossover Module - CM}

The crossover module detailed here in this section implements single point crossover. The architecture proposed here contains $\frac{N}{2}$ crossover modules and each one consists of four bit splitters, two identical crossover submodules, and two concatenators. Similarly to the FFM described in Section \ref {FFModule_Section}, the CM also has chromosome splitters in order to manipulate the two variables stored in $w[m](k)$ independently.

As seen in Figure \ref{Cruzamento_esquema}, the two input chromosomes, $w_{j-1}[m](k)$ and $w_j[m](k)$, are split into two halves, each. The first $w_{j-1}[m](k)$ half is sectioned by the splitter CMDIV1$_j $ which is renamed $pw_{j-1}[\frac{m}{2}](k)$ and the second half of that same variable is sectioned by the splitter CMDIV2$_j $ and becomes  $qw_{j-1}[\frac{m}{2}](k)$. The same happens with the chromosome $w_{j}[m](k)$ which is sectioned into $pw_{j}[\frac{m}{2}](k)$ and $qw_{j}[\frac{m}{2}](k)$ through the divisors CMDIV3$_j$ and CMDIV4$_j$, respectively.

Separating the variables of each chromosome, they are forwarded to the CM submodules CMPQ1$_j $ and CMPQ2$_j $ so then the crossing is performed. This is conducted in such a way that the crossover is performed between similar variables, that is, the first variable $pw_{j-1}[\frac{m}{2}](k)$ from the chromosome $w_{j-1}[m](k)$ will be crossed with the first variable $pw_{j}[\frac{m}{2}](k)$ from the chromosome $w_{j}[m](k)$. 

In the case of single variable problems the system works in an equivalent way. Only the least significant half of the variables $w_{j-1}[m](k)$ and $w_j[m](k)$ will contain useful data and only block CMPQ2$_j $ will handle nonzero data.

\begin{figure*}[t]
\begin{center}
  \includegraphics[scale=0.65]{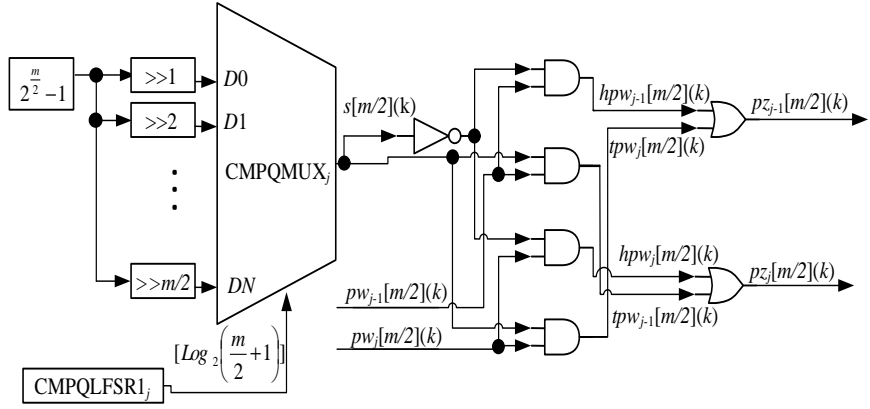}\\
  \caption{$J$-it Crossover Module (CMPQ1$_j$)}
  \label{Cruzamento_esquema_pq}
\end{center}
\end{figure*}

Figure \ref {Cruzamento_esquema_pq} presents in detail the circuit of the $j$-th crossover submodule named CMPQ1$_j $. It is composed by a $\frac{m}{2}$-input MUX, called here CMPQMUX$_j $, whose purpose is to randomly select one of the $\frac{m}{2}$ possible cutting points. The selection of each CMPQMUX$_j $ is controlled by the pseudo random number generator CMPQLFSR1$_j $ whose output signal, $\text{CMPQ}r_{1i}[32](k)$, is truncated in the $ \left \lceil log_2(\frac{m}{2}+1) \right \rceil$ more significant bits before entering the MUX selector.

The selection of the CMPQ1$_j $ cut-off point is done relying on the mask originated from the constant $2^{\frac{m}{2}}-1$. This constant creates a vector of $1$s of the size of the chromosome to be crossed, in the case $\frac{m}{2}$. Then, a random and zero-padding right shift is performed according to CMPQLFSR1$_j $ value. This displacement will transform the vector of $1$s into a vector of $0$s and $1$s concatenated and still of size $\frac{m}{2}$. This mask and its inverse will be responsible for carrying out the crossover operation aided by the AND and OR logic gates shown also in Figure \ref {Cruzamento_esquema_pq}.

Equations \ref {mascara2} and \ref {mascara3} exemplify a case where $ m = 20$ and CMPQMUX$_j$ shifts the value of $2^{\frac{m}{2}}-1$ three times
\begin{equation} \label{mascara1}
2^{\frac{m}{2}}-1 = 1111111111;
\end{equation}
\begin{equation} \label{mascara2}
s_i\left[\frac{m}{2}\right](k) = 0001111111;
\end{equation}
\begin{equation} \label{mascara3}
\neg s_i\left[\frac{m}{2}\right](k) = 1110000000.
\end{equation}

In each $k$-th generation, the two entries of the module CMPQ1$_j $, the variables $pw_{j-1}[\frac{m}{2}](k)$ and $pw_{j}[\frac{m}{2}](k)$ are divided into head
\begin{equation} \label{EqMC1}
hpw_{j-1}\left[\frac{m}{2}\right](k) = \neg s_j\left[\frac{m}{2}\right](k) \land pw_{j-1}\left[\frac{m}{2}\right](k)
\end{equation} 
\begin{equation} \label{EqMC2}
hpw_{j}\left[\frac{m}{2}\right](k) =  \neg s_j\left[\frac{m}{2}\right](k) \land pw_{j}\left[\frac{m}{2}\right](k)
\end{equation} 
and tail
\begin{equation} \label{EqMC3}
tpw_{j-1}\left[\frac{m}{2}\right](k) =  s_j\left[\frac{m}{2}\right](k) \land pw_{j-1}\left[\frac{m}{2}\right](k),
\end{equation}
\begin{equation} \label{EqMC4}
tpw_{j}\left[\frac{m}{2}\right](k) = s_j\left[\frac{m}{2}\right](k) \land pw_{j}\left[\frac{m}{2}\right](k),
\end{equation}
where $s[\frac{m}{2}](k)$ is the CMPQMUX output. After this step, the crossover will be performed by concatenating the head of parent 1, $hpw_{j-1}[\frac{m}{2}](k)$, with the parent's tail 2, $tpw_{j}[\frac{m}{2}](k)$, and the parent head 2, $hpw_{j}[\frac{m}{2}](k)$, with the parent's tail 1, $tpw_{j-1}[\frac{m}{2}](k)$, thus giving rise to two new chromosomes of the new population,
\begin{equation} \label{EqMC21}
pz_{j-1}\left[\frac{m}{2}\right](k) = hpw_{j-1}\left[\frac{m}{2}\right](k) \lor tpw_{j}\left[\frac{m}{2}\right](k).
\end{equation}
and
\begin{equation} \label{EqMC26543}
pz_{j}\left[\frac{m}{2}\right](k) = hpw_{j}\left[\frac{m}{2}\right](k) \lor tpw_{j-1}\left[\frac{m}{2}\right](k).
\end{equation}

For the CMPQ2$_j $ submodule the equivalent happens. In this case, the input values will be $qw_{j-1}[\frac{m}{2}](k)$ and $qw_{j}[\frac{m}{2}](k)$ and the outputs will be $qz_{j-1}[\frac{m}{2}](k)$ and  $qz_{j}[\frac{m}{2}](k)$.

After the similar variables have been crossed within each submodule, they are directed to the outputs of each respective CMPQs where the concatenators CMCCAT1$_j$ and CMCCAT2$_j$ will give rise to new individuals (chromosomes) from the population by concatenating both the parts forming them, $pz[\frac{m}{2}](k)$ and $qz[\frac{m}{2}](k)$ (Figure \ref{Cruzamento_esquema}).

It is important to emphasize that after $\frac{N}{2}$ CMs have performed their operations, $ N $ new chromosomes that will form a new population will have been created. Some of these individuals will pass through the MM (to be described in Section \ref {MM_label}) before the start of the next generation, but always at the end of each iteration, $ N $ new individuals will have been created so that the GA population will always remain with $ N $ chromosomes.

\subsection{Mutation Module - MM}
\label{MM_label}

\begin{figure}[h]
  \centering
  \includegraphics[width=6cm,keepaspectratio]{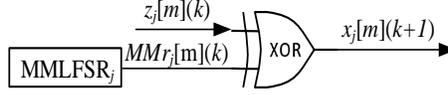}
  \caption{Mutation Module - MM}
  \label{MM_fig}
\end{figure}

As in the Algorithm \ref {ResumoAG} in Line \ref {CodFM} the mutation operation will be performed on a group of $ P $ individuals, that is, there are $ P $ mutation modules and each $j$-th module, MM$_j$, changes the value of the chromosome to be mutated through an XOR operation with a number created randomly by an associated generator called MMLFSR$_j$ (Figure \ref {MM_fig}). The $ P $ MM will modify the first $ P $ individuals of the population as shown in Figure \ref {AG_Estrutura_Geral}.

The output of the $j$-th, MM$_j$, module in every $k$-th generation can be expressed by
\begin{align}\label{EqMC432}
x_j[m](k+1) &= (\neg z_j[m](k)  \land \text{MM}r_j[m](k))   \nonumber \\
&\qquad \lor  (z_j[m](k)  \land \neg \text{MM}r_j[m](k)) 
\end{align}
where $\text{MMr}_{j}[m](k)$ represents the pseudo random number generated by $j$-th MMLFSR$j$.

In the case of single-variable problem optimization, this mutation operation will possibly assign non-zero values to the $\frac{m}{2}$ unused bits of the mutated chromosome. However, this will not be a problem since these $\frac{m}{2}$ bits will be zeroed when passing through the FFM in the following generation.

\subsection{Syncronization Module - SyncM}
\label{Section_Sync}

Finally, the last module is the synchronization module. It aims to enable the registers, responsible for storing the population chromosomes of the genetic algorithm, to receive new values. These new values result from the mutation and crossover processes of the previous generation and are stored in the RX registers to initiate a new iteration of the algorithm.

\begin{figure}[h]
  \centering
  \includegraphics[width=6cm,keepaspectratio]{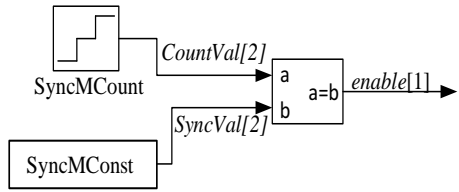}
  \caption{Syncronization Module - SyncM}
  \label{SyncM_fig}
\end{figure}

This module contains a counter, a constant value and a comparator as shown in Figure \ref{SyncM_fig}. The variable $enable$ is enabled when the comparison returns a true value, that is, when the counter value matches the value stored in the constant. The $SyncVal [2] $ value is obtained according to the implemented design, and it is adjusted according to the delay that the system needs to perform all its operations and provide a new set of chromosomes. The output value of this module is a boolean value, and the values of the counter and constant output are 2-bit values. This number was chosen because it was the maximum delay found in the implementation for an entire generation, a delay for each ROM of the FFM.

In all the tests performed in this work the GA operations were performed at a sampling rate
\begin{equation}
\label{Rg_label}
R_g=\frac{3}{T_g}
\end{equation}
where $T_g$ is the time for each $k$-th generation be finished.

Although $ R_g $ is the maximum possible sample rate to operate the system and $T_g$ is the minimum equivalent time, the equation \ref{Rg_label} divides these values by 3 since only every three clocks a new population is originated in the GA, since there are two delays in the architecture between the beginning of the $k$ generation and the end of it. Thus, these two delays caused by the LUTs contained in the FFM (Section \ref {FFModule_Section}) make the frequency $\frac{3}{T_g}$ the one which will process the population $k + 1$ from an earlier population $k$.

Generally, if the architecture contained any $\eta$ components that caused system delays, the sample rate $R_{g\eta}$ of this system would be
\begin{equation}
\label{Rg_label_generico}
R_{g\eta}=\frac{\eta}{T_{g\eta}}.
\end{equation}

\section{Results}\label{Results_sec}

Aiming to validate the proposed implementation of the GA on FPGA, simulations, analyses and syntheses were performed in the optimisation of different functions for various population sizes. The first function, called here F1, used in the tests to validate the proposal was an one variable function expressed as
\begin{equation}
\label{F1}
f(x) = x^3 -15x^2 + 500,
\end{equation}
The second function, called here F2, was
\begin{equation}
\label{F2}
f(x,y) = 8x - 4y + 1020,
\end{equation}
and, lastly, the last function, here called F3, was the function
\begin{equation}
\label{F3}
f(x,y) = \sqrt{x^2+y^2}.
\end{equation}

This work has implemented and synthesised on FPGA the three functions previously mentioned for populations of size $ N  = 4$, $ N  = 8$, $ N  = 16$, $ N  = 32$ and $ N  = 64$ and for chromosomes with size $ m  = 20$, $ m  = 22$, $ m  = 24$, $ m  = 26$, and $ m  = 28$ bits.

It is important to emphasise that these functions were chosen for comparison reasons, since they have already been used in the state of the art in previous works that will be shown next. However, the implementation proposed is capable of implementing any function in the format shown by Equation \ref{EqGenerica} requiring only the modification of the values stored in the memories.

All results were obtained using the development platform and a FPGA Virtex 7 xc7vx550t-1ffg1158. The Virtex 7 FPGA used has $ 86,600 $ slices that group $ 692,800 $ flip-flops, $ 554,240 $ logical cells that can be used to implement logical functions or memories and $ 2,880$ DSP cells with multipliers and accumulators.

\begin{figure}[h]
\begin{center}
  \includegraphics[scale=0.35,keepaspectratio]{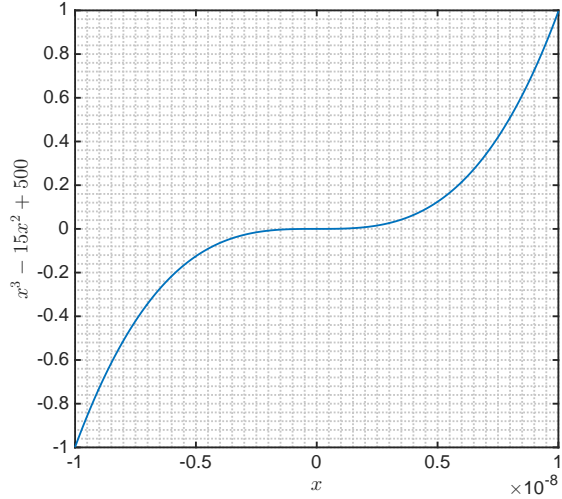}\\
  \caption{Fitness Function F1: \(f(x) = x^3 -15x^2 + 50\).}
  \label{Plot_F1}
\end{center}
\end{figure}

As previously mentioned, three different functions were minimised for the validation of the implementation. The first one was the function F1 presented in Equation \ref{F1} and shown in Figure \ref{Plot_F1}.

\begin{figure}[h]
\begin{center}
  \includegraphics[scale=0.4,keepaspectratio]{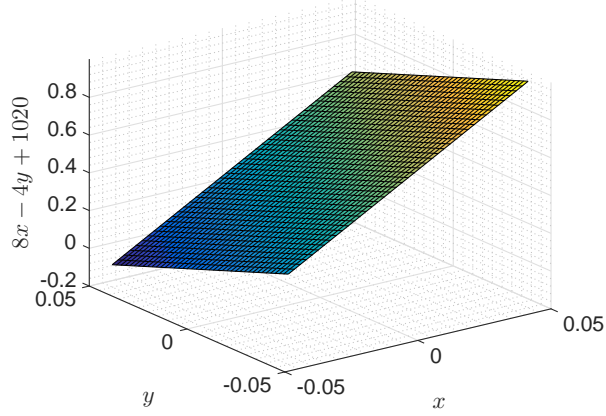}\\
  \caption{Fitness Function F2: \(f(x,y) = 8x - 4y + 1020\)}
  \label{Plot_F2}
\end{center}
\end{figure}

This function was chosen because it was previously used by \citep{[14]} to validate its proposal which developed a high-speed Genetic Algorithm on FPGA. Regarding the implementation of this function as described in Equation \ref{EqGenerica}, the following associations can be made:
\begin{equation} \label{F1a}
\alpha(px) = 0,
\end{equation}
\begin{equation} \label{F1b}
\beta(qx) = (qx)^3 -15(qx)^2 + 50,
\end{equation}
\begin{equation} \label{F1c}
\gamma(\delta) = \delta,
\end{equation}
Therefore, F1 can be represented by
\begin{equation} \label{F1d}
y(px,qx) = 1*((qx)^3 -15(qx)^2 + 50+ 0).
\end{equation}
The second function is presented in Figure \ref{Plot_F2} and has been previously used in \citep{[5]} to validate the implementation of a customisable GA IP core for general purposes.

Regarding the implementation of F2 as described in Equation \ref {EqGenerica}, the following associations can be made:
\begin{equation} \label{F2a}
\alpha(px) = 8(px),
\end{equation}
\begin{equation} \label{F2b}
\beta(qx) = -4(qx) + 1020,
\end{equation}
\begin{equation} \label{F2c}
\gamma(\delta) = \delta,
\end{equation}
Therefore, F2 can be represented by
\begin{equation} \label{F2d}
y(px,qx) = 1*(8(px) -4(qx) + 1020).
\end{equation}

\begin{figure}[h]
\begin{center}
  \includegraphics[scale=0.4,keepaspectratio]{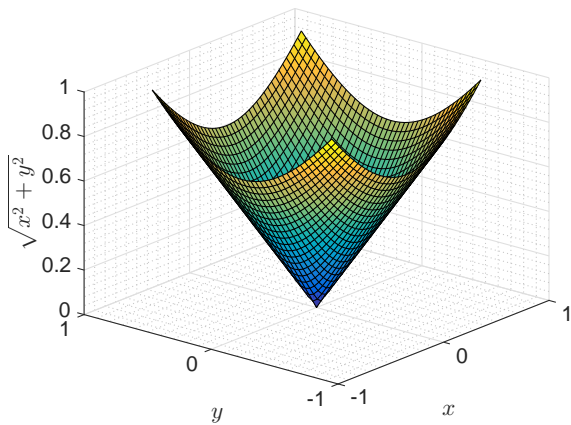}\\
  \caption{Fitness Function F3: \(f(x,y) = \sqrt{x^2+y^2}\)}
  \label{Plot_F3}
\end{center}
\end{figure}

Finally, Figure\ref {Plot_F3} shows the last function used to validate the proposal presented here. This function could be seen previously with the same use in \citep{[3]} and \citep{[17]}. Both works use GA, but only \citep{[3]} implements the algorithm on FPGA.

Similarly, the F3 in the parameters of the equation \ref{EqGenerica}, can be seen as follows:
\begin{equation} \label{F3a}
\alpha(px) = (px)^2,
\end{equation}
\begin{equation} \label{F3b}
\beta(qx) = (qx)^2,
\end{equation}
\begin{equation} \label{F3c}
\gamma(\delta) = \sqrt\delta,
\end{equation}
Therefore, F3 can be represented by
\begin{equation} \label{F3d}
y(px,qx) = \sqrt{(px)^2+(qx)^2}.
\end{equation}

The parameters used in the experiments were based on configurations of previous experiments found in the literature together with some empirically obtained configurations. For the number of generations $ k $, it was experimentally observed that for all evaluated functions, the minimum value sought was obtained before the $ 100 $ GA generations were reached. This number is in agreement with what was found in the literature, as can be seen in \citep{[14]}, for example. Thus, $ k = 100 $ was adopted as default value for the optimization experiments performed here.

Similarly, the GA population sizes to be implemented and synthesised in the FPGA were determined. As seen in \citep{[50]} the population size was $ N  = 32$. In \citep{[30]} the population had size $ N  = 16$ and in \citep{[51]} GA was implemented with populations of sizes $ N  = 64$ and $ N  = 128$. Thus, the architecture of the proposal presented here was implemented for the five population sizes already mentioned before. The aim behind these different sizes of $ N $ was to compare how much this parameter influences the convergence, speed and area occupation in the FPGA.

\begin{figure}[h]
\begin{center}
  \includegraphics[scale=0.33,keepaspectratio]{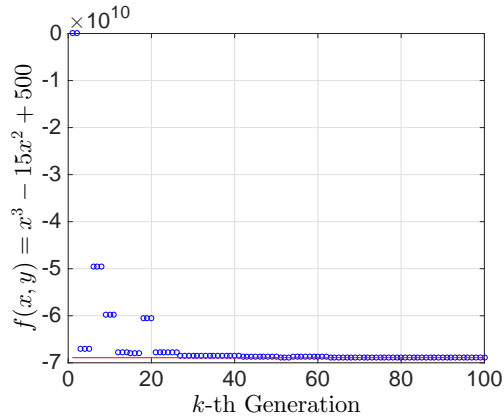}\\
  \caption{Optimising F1.}
  \label{F1_min}
\end{center}
\end{figure}

\begin{figure}[h]
\begin{center}
  \includegraphics[scale=0.33,keepaspectratio]{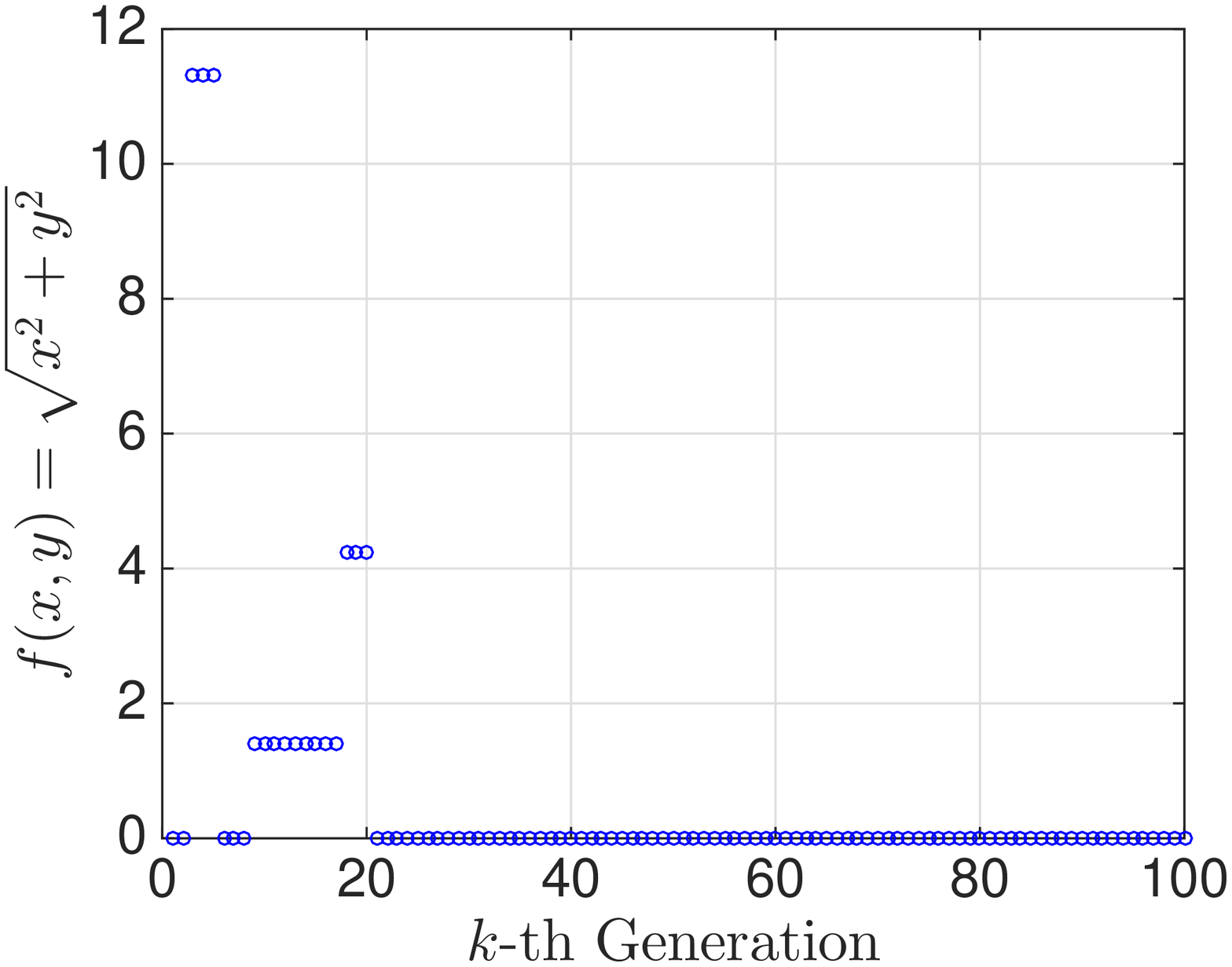}\\
  \caption{Optimising F3.}
  \label{F3_min}
\end{center}
\end{figure}

Finally, for the same purpose of comparing how much a parameter influences certain convergence and synthesis characteristics, the bit length  $ m $  varied for all population sizes as also quoted previously. The Figures \ref {F1_min} and \ref {F3_min} picture the operation of the proposed GA  for the fitness functions F1 and F3, respectively.

The fitness function 1 (Equation \ref {F1}) was minimized using the GA with a population size of $N = 32$ and $m = 26$. Thus, $[\frac{m}{2}]$ bits were used for each variable from the fitness function. Given that this is a single-variable problem, the function made use of only $[\frac{m}{2}]$ bits. For the minimization shown in Figure \ref {F1_min} the range of the F1  was of $f(-2^{12})$ to $f(2^{12}-1)$. Thus, the minimum possible value in the range is $f(-2^{12}) = -6.8971*10^{10}$. As depicted, it is noticed that the global minimum was reached approximately in the half of the $ 100 $ generations, thus proving the functionality of the proposed system.

Similarly, the fitness function 3 (Equation \ref {F3}) was also minimised, as shown in Figure \ref {F3_min} but with a population size of $N  = 64$ and $ m =  20$. In this case, $ f (x, y) $ only allows results greater than or equal to zero when working in the domain of real numbers, so the smallest possible value is zero. The parallel Genetic Algorithm implemented on FPGA proposed herein has managed to minimize F3 in a little over $ 20$ iterations of the algorithm. This is not a fixed value, since the GA is a stochastic algorithm, but with this value it is possible to have an idea of the number of generations required for convergence.

Both results were obtained from the average of multiple results. It is also important to emphasise that parameters such as the range of values to be calculated, bit width ($ m$), decimal precision and the possibility of exploring negative numbers are all parameters of the LUT (Section \ref{FFModule_Section}) and configurable by the user. As already mentioned, the option of maximising or minimising the function to be optimised is also another configurable variable.

The Table \ref {Resultados_da_sintese_do_algoritmo_genetico_na_FPGA} presents the synthesis results in the target FPGA for various population sizes and $ m $ = 20. It is clear in all scenarios that the area of occupation, clock consequently the number of generations per second, $ R_g $, are parameters considerably sensitive to the population size, $N$. Here, the $ R_g $ represents the number of generations performed in the GA per second, which can also be interpreted as some possible solutions which the system provides in that interval. Equation \ref{Rg_label} states that this number is equal to $\frac{3}{Tg}$, that is, the inverse of the time of each GA generation divided by three. This is explained because the system generates two delays when placing two ROM memories in series in the FFM described in Section \ref {FFModule_Section}. Consequently, a new GA population is generated only after three system clocks.

The clock is the maximum frequency the system performs when implementing this architecture, and it does not take into account the delays required to generate a new population. The clock represents only the hardware speed for that specific implementation, so it is $3\times$ faster than the number of generations performed in the GA per second.

\begin{table}[h]
    \begin{center}
    \caption{GA synthesis on FPGA for $ m $ = 20.}
        \begin{tabular}{|c|c|c|c|c|}
         \hline
             $N$ &    Registers & Logic Cells & Clock & Generations\\
                  &       Flip-flops    &        (LUTs)        &      (MHz)   & Per Second $\times 1000$ \\
             \hline
             $4$ & $457$ & $592(1\%)$ & $50.28$ & $16.76$    \\
              \hline
             $8$ & $839$ & $1.558(1\%)$ & $49.32$ & $16.44$   \\
            \hline
             $16$ & $1.616$ &$4,400(1\%)$&	$49.32$ & $16.44$  \\
            \hline
              $32$ & $3.225$ &$15,908(4\%)$ &	$48.51$ & $16.17$  \\
            \hline
                $64$ & $6.598$&	$58.875(16\%)$&	$34.56$ & $11.52$  \\
            \hline
        \end{tabular}
        \label{Resultados_da_sintese_do_algoritmo_genetico_na_FPGA}
    \end{center}
\end{table}

\begin{figure}[h]
\begin{center}
  \includegraphics[scale=0.33,keepaspectratio]{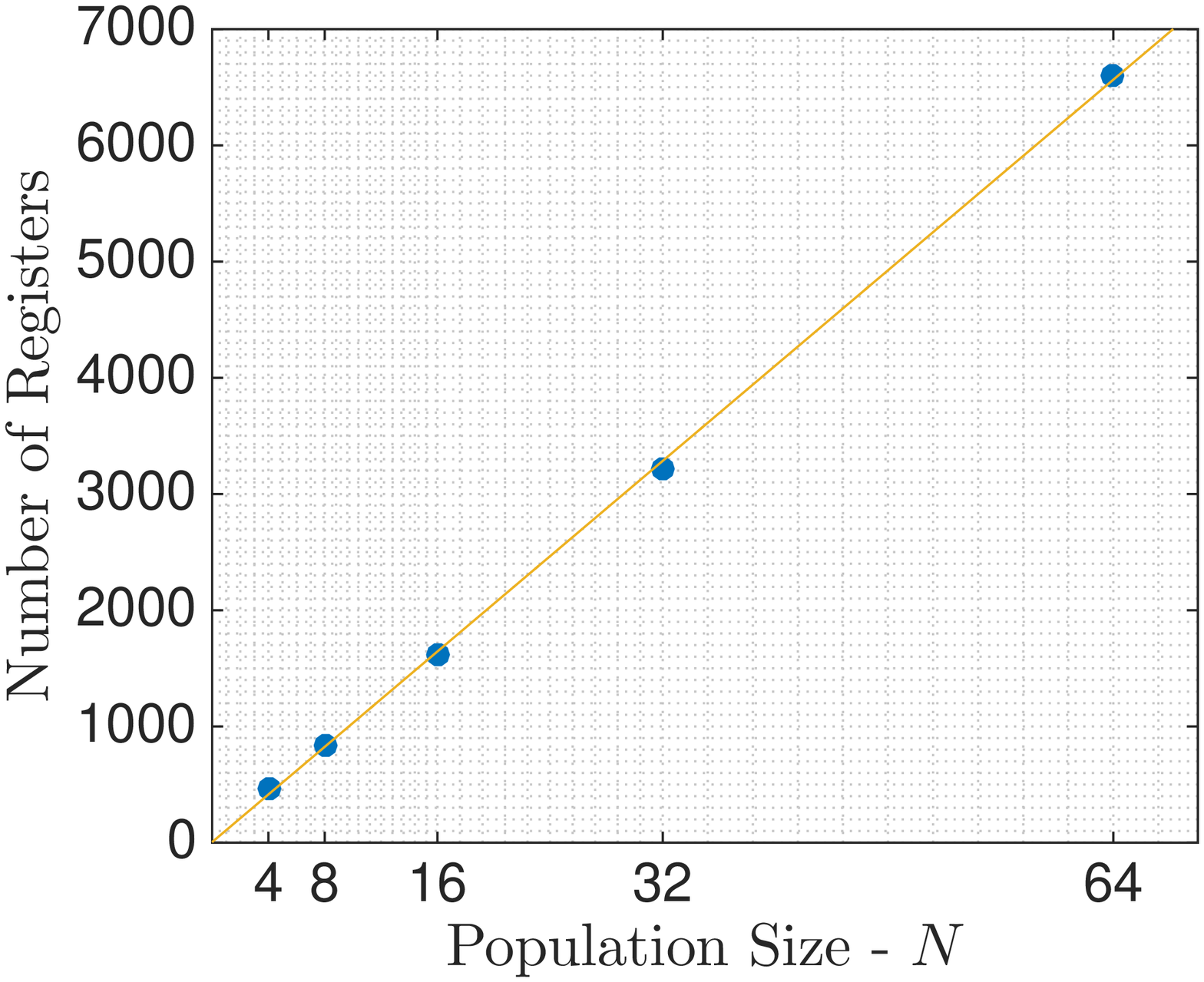}\\
  \caption{Registers' occupation in the FPGA varying with $ N $.}
  \label{Crescimento_Reg}
\end{center}
\end{figure}

The area occupation spent in registers (Figure \ref {Crescimento_Reg}), presented in the second column of the Table \ref {Resultados_da_sintese_do_algoritmo_genetico_na_FPGA}, is due to the storage of population values in RX (Figure \ref {AG_Estrutura_Geral}) and the  pseudo random number generators, mainly. This occupation increases linearly according to $ N $, since the larger the population, the greater the number of RX registers required, as well as operations that require the pseudo random number generators. Figure \ref {Crescimento_Reg} shows this growth graphically with a linear interpolation.

\begin{figure}[h]
\begin{center}
  \includegraphics[scale=0.33,keepaspectratio]{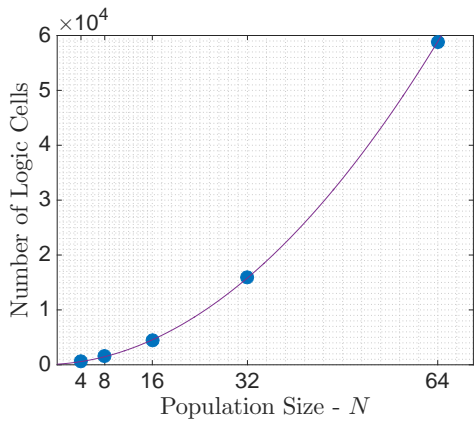}\\
  \caption{FPGA LUTs occupancy varying with $ N $}
  \label{Crescimento_LUTs}
\end{center}
\end{figure}

The logical cells (LUTs) occupation, presented in the third column of the Table \ref {Resultados_da_sintese_do_algoritmo_genetico_na_FPGA}, was increasing and not linear with $ N $, as can be seen in Figure \ref {Crescimento_LUTs}. This nonlinear growth is caused by the selection module (Subsection \ref {SM_Section}) that for each $j$-th module, SM$_j$, there are three  $ N $ inputs multiplexers ( SMMUX1$_j$, SMMUX2$_j$ and SMMUX3$_j$ ). 

According to  \citep{[48]}, each Virtex 7 logical cell can construct four 1-input MUXs, thus to build a a $N$-inputs multiplexers , approximately $\frac{N}{4}$ logical cells are required, totalling approximately $\frac{3N}{4}$ cells for each SM$_j$ (SMMUX4$_j$, SMMUX5$_j$ and SMMUX6$_j$) have not been considered). Since there are $ N $ SM modules, there are approximately $\frac{3N^2}{4}$ logical cells for each bit of the input bus of the MUX. Thus, the exponential growth result from the use of the logical cells is explained.

In this context, it is important to note that implementation with $ N = 64 $ individuals does not reach even one fifth of the FPGA cells (around $16\%$ of Virtex 7). This is a positive indicator for implementations with larger populations.

Finally regarding the table, the last two columns show the Clocks and the number of generations performed in the GA per second for each value of $ N $ and there is speed reduction according to the population growth. Theoretically, if all the modules were independent (specific for each individual) this reduction should not happen, however, it is observed that in the selection modules, SM$_j$ (Figure \ref {SM_Section}), there is a dependency between the $ N $ chromosomes (due to information sharing) causing a join in the circuit and thus, an increase in processing time. On the other hand, it is also observed that the reduction rate is not linear, which favours the implementation. Another important information to note is that even with the reduction, each GA generation of $ 64 $ chromosomes is generated in $Tg\approx87 \, \text{ns}$, in other words $ 87 $ millions of generations to every $ 1 $ ms. This result has a very significant impact and makes the use of GA possible in several real-time embedded applications such as robotics, telecommunications and others.

\begin{figure}[h]
\begin{center}
  \includegraphics[scale=0.33,keepaspectratio]{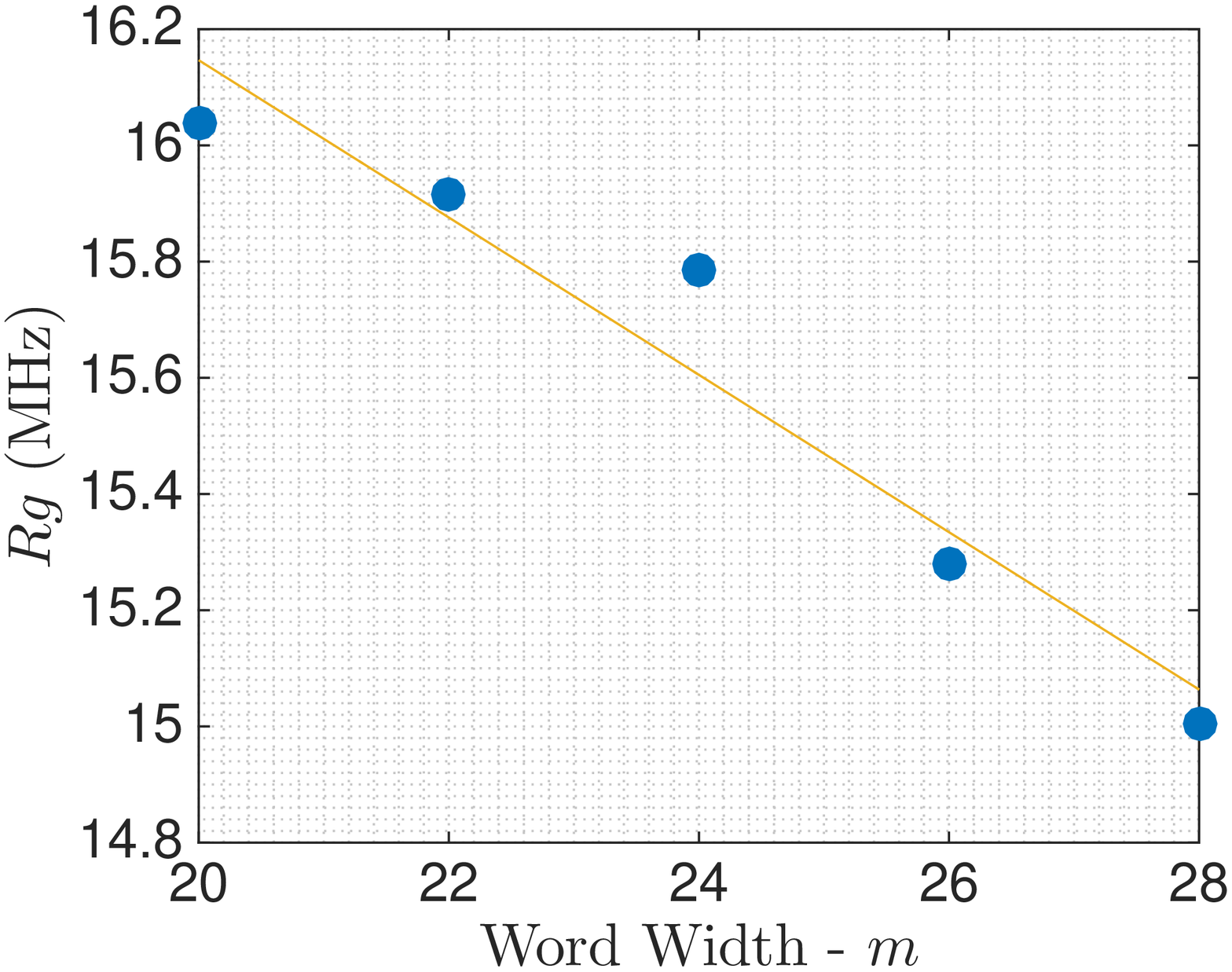}\\
  \caption{Decrease of $Rg$ when varying $m$.}
  \label{Rg_por_m}
\end{center}
\end{figure}

The Figure \ref{Rg_por_m} represents the influence of the bit width $ m $ on the Clock for a GA with $N  = 32$. It is noted a decrease of the processing speed with the increase of the number of bits, however this fall is not significant. The clock variation is only slightly more than $1$ MHz when the implementation is compared using $ m = 20$ with the implementation using $m = 28$. The interpolation shown in the Figure suggests a linear fall .

\begin{figure}[h]
\begin{center}
  \includegraphics[scale=0.33,keepaspectratio]{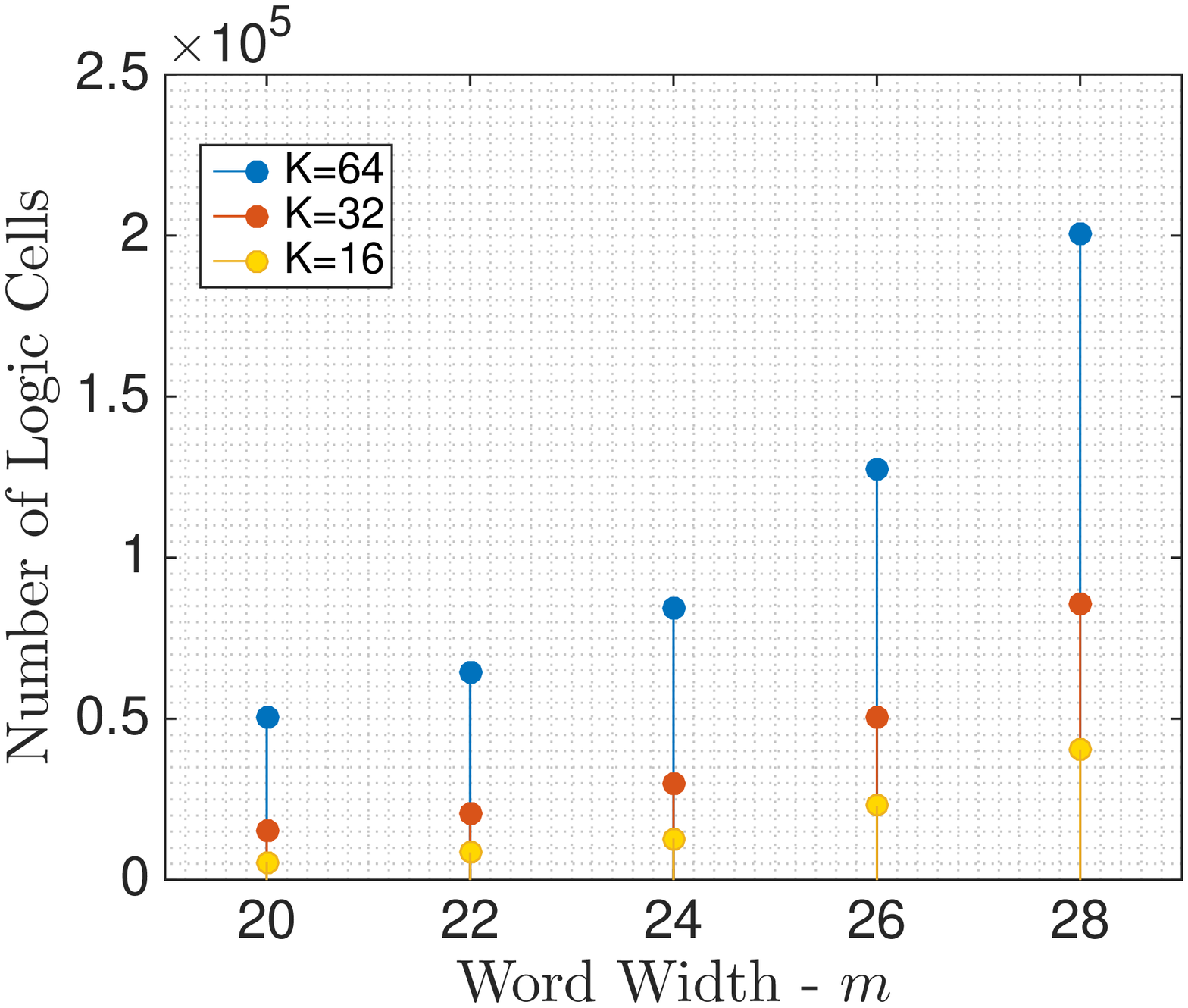}\\
  \caption{Relation between the LUTs usage with the increase of $ m $.}
  \label{d}
\end{center}
\end{figure}

The last Figure (\ref {d}) illustrates the relationship between the increase of LUTs used in the FPGA and the variation of the bit width  $ m$ for three different population sizes. A larger difference is observed between the quantities of LUTs used in $m = 28$, mainly due to the nonlinear growth of these components comparing to $ N $. However, as already seen in Figure \ref {Rg_por_m} the increase of $m$ is also a factor responsible for slowing the processing speed, $ Rg $.

Analyzing the synthesis results, it was noticed that different fitness functions such as F1, F2 and F3 did not result in significant differences in the LUTs consumption and Registers in the FPGA, as well as no significant differences were observed in $ Rg $. This result was already expected, since the only variation that occurs when changing the fitness function is the content of the FFM LUTs. Thus, it is possible to extend this thinking and assert that the values of the Table \ref {Resultados_da_sintese_do_algoritmo_genetico_na_FPGA} are true for any other function, in the parameters of Equation \ref {EqGenerica}, using $m= 20$ bits.

\section{Comparisons with state of the art works}
\label{Comparisons_sec}

Following, comparisons of the results obtained by the proposed implementation with equivalent results found in works belonging to the state of the art are presented. The comparisons which will be shown below and which are summarised in the Table \ref {tabela_comparacao} were made with the greatest similarity of parameters as possible. The table presents a column that presents the comparative references, the next two columns show the parameters of the GAs compared, then the times obtained by the works of the state of the art are shown and, finally, the results obtained by the implementation presented here and the respective speedups are displayed.

The system presented by \citep{[14]}, a high-speed implementation of GA on FPGA, demonstrated a runtime of $ 0.21 $ milliseconds for a GA implemented on FPGA with $k =100$ generations and a population of size $N= 32$. For the same settings, the system proposed here achieved a time of$\approx6.18$ microseconds, which proves to be $\approx34 \times$ faster.

Similarly, the implementation presented by \citep{[51]} also presented an GA on FPGA with population size$ N  = 32$, chromosome size of $ m = 16$ and $ k  = 60$ generations. The implementation validated its proposal with the traveling salesman problem and resulted in a running time of $1.702m$s. Although a test in the same parameters of \citep{[51]} has not been performed here, a comparison can still be made due to the versatility of problems solved by different LUT as shown in the FFM. An AG with $ k  = 60$, $ N  = 32$ and $ m  = 20$ can be solved in  $\approx3.71$ microseconds in the work presented here, meaning a time $\approx459 \times$ faster than in \citep{[51]}.

In similar fashion, the work of \citep{[5]} presented a highly programmable GA IP core on FPGA. For a setting of $ k  = 32$ and $ N  = 32$ the authors stated a speedup of $5.16 \times$ over an equivalent software implementation which achieved a running time of $ 37.615 $ milliseconds. In a comparison, the implementation presented here performed the equivalent situation in a time of $\approx1.98$ microseconds, which represents a speedup of $\approx19007 \times$ over the serial implementation shown in \citep{[5]}. In other words, a time $\approx3683 \times$ less than the GA IP core proposed by \citep{[5]}.

Finally, the implementation here proposed can also be compared to the work published in \citep{[20]}. As already mentioned in Section \ref {Related_work_section}, this article presents the OIMGA, an implementation of a monogenetic FPGA algorithm that retains only the best chromosome of the generation. In one of the tests performed to validate the implementation, the authors optimised a one variable function with a population of $ N  = 64$ in $\approx0.8$ seconds. In a scenario where the proposed parallel GA take this time to solve the same function it would process $ k =  9.2$ million of generations. Of course, this value is unreasonable for that function. As shown previously in the results, $ k =  100$ generations was the default value to optimise functions of one or two variables, thus, even if the number of generations needed to optimise the same function was $ k =  500$ (a generous estimate), the time resulting from the implementation proposed by \citep{[20]} would still be $\approx18432 \times$ higher.

\begin{table*}[h]
\centering
\caption{Comparative table with state of the art works}
\label{tabela_comparacao}
\begin{tabular}{|c|c|c|c|c|c|}
\hline
Reference &   N & k   & \begin{tabular}[c]{@{}c@{}}Reference\\ Time\end{tabular} & \begin{tabular}[c]{@{}c@{}}Obtained \\ Time\end{tabular} & Speedup \\ \hline
\citep{[14]}     &  32 & 100 & $0.21$ ms         & $6.18\mu$s   & 34                \\ \hline
\citep{[51]}     &  32 & 60  & $1.702$ ms        & $3.71\mu$s   & 459               \\ \hline
\citep{[5]}      &  32 & 32  & $7.29$ ms         & $1.98\mu$s   & 3683              \\ \hline
\citep{[20]}     &   64 & 500 & $0.8$ s             & $43.40\mu$s  & 18432             \\ \hline
\end{tabular}
\end{table*}

\section{Conclusion}\label{conclusion_sec}

After the presentation of the results in Section \ref {Results_sec} it can be affirmed that the implementation proposal was in fact validated and fulfilled with its objective of being a parallel implementation of high-performance of a GA. The synthesis results confirmed that the present proposed parallel implementation of AG on FPGA is able to optimize a wide range of functions in a viable time for critical applications that require short time constraints or a large amount of data to be processed in a short interval.

Comparisons with other implementations found in the literature in Section \ref {Comparisons_sec} reinforce the high speed achieved by the implementation developed. This enables the use of this system in a commercial context for applications such as Internet Touch, robotics, real-time applications and medical applications. In addition, this system has proven to be an acceleration tool for any hardware system that makes use of genetic algorithms.

As well as the high-performance achieved, the small area consumption of the implementation developed here is a notorious feature. This makes it possible for other systems to also be embedded in the FPGA, since the on-board GA occupies less than $\frac{1}{5}$ of the Virtex 7 logic cells used as a test. This logical cells low consumption feature is essential for applications where the area is the biggest constraint as spatial applications, for example.

The experiments carried out proved that the sizes of $ N $ tested are sufficient to solve most of the practical problems as the literature says. It has been found that the duration in iterations ($ k $) of GA does not need to be greater than a few hundred. It has been proven that a few hundred generations or even $ k  = 100$ is a reasonable number of generations for a GA. The parameter $ m $ proved to be of great importance, since it directly affects the GA convergence speed, the area occupied on the FPGA, the response precision, as well as the achieved $R_g$.

\bibliography{Paper14Main}

\end{document}